\documentclass[aps,amsmath,amssymb,reprint,twocolumn,prl,superscriptaddress,notitlepage,showpacs,tightenlines]{revtex4-1}
\usepackage{exscale}
\usepackage{relsize}
\usepackage{graphicx}
\usepackage{dcolumn}
\usepackage{bm}
\usepackage{textcomp}
\usepackage{siunitx}
\usepackage{appendix}
\usepackage{array}
\usepackage{color}
\usepackage{textcomp}
\usepackage{amsmath}
\usepackage{float}
\usepackage{verbatim}
\usepackage[colorlinks,citecolor=blue,linkcolor=red,hyperindex,CJKbookmarks]{hyperref}

\allowdisplaybreaks[3]

\begin{document}

\title{Correlation Imaging via Hybrid Entanglement between Microwave Photons and Surface Acoustic Wave Phonons}

\author{Yu-Yuan Chen}
\affiliation{School of Integrated Circuits, Tsinghua University, Beijing 100084, China}

\author{Ling-An Wu}
\affiliation{Institute of Physics, Chinese Academy of Sciences, Beijing 100190, China}

\author{Yu-xi Liu}
\email{yuxiliu@mail.tsinghua.edu.cn}
\affiliation{School of Integrated Circuits, Tsinghua University, Beijing 100084, China}
\date{\today}

\begin{abstract}
Establishing effective spatial correlations between distinct quantum fields is essential for multi-physics quantum sensing.~We here propose a correlation imaging approach based on the hybrid entanglement between microwave photons and microwave surface acoustic wave phonons generated via a superconducting quantum circuit.~At matched microwave frequencies, the five order-of-magnitude wavelength disparity between photons and phonons leads to a unique quantum entanglement resource, which allows the optimal field for probing an object to be distinct from that used for readout.~This resource enables two imaging modalities with giant spatial scaling: a $10^{-8}-10^{-6}$-fold demagnification of macroscopic object profiles onto microscopic phononic chips, and conversely, a $10^{4}-10^{6}$-fold magnification of microscopic object profiles into macroscopic photonic readout devices. By interfacing free-space long-wavelength microwaves with on-chip short-wavelength surface acoustic waves, our correlation imaging approach provides a general framework for cross-scale quantum sensing.
\end{abstract}
\maketitle

{\it {Introduction.---}}
Hybrid quantum architectures, including either interfacing distinct quantum platforms via mediating quantum fields~\cite{Hybrid-RMP-Xiang,Hybrid-QD-Mech-Yeo,Hybrid-NV-Mech-Arcizet,SAW-Qcontrol-Satzinger,SAW-Qcontrol-Dumur} or integrating distinct quantum fields via a common quantum system~\cite{SAW-Propagating-Gustafsson,SAW-cQAD-Manenti,SAW-Transducer-Schuetz,SAW-Superconducting-MW-Zeng}, have shown great potential in quantum information processing and multi-physics quantum sensing. Recent studies show that superconducting quantum circuits~\cite{Hybrid-RMP-Xiang,Superconducting-Circuit1,Superconducting-Circuit2}~can be coupled to microwave photons and microwave surface acoustic wave phonons~\cite{QuantumAcoustics-Multimode-AW,SAW-Cavity-Multimode-TunableCoupling,SAW-Piezomagnetism,SAW-Skyrmion,
QuantumAcoustics-Article1,SAW-Entanglement1}~simultaneously or separately. Hereafter, for convenience, we use photons and phonons to denote microwave photons and surface acoustic wave phonons. It is well known that phonons propagate about five orders of magnitude slower than photons~\cite{SAW-Background-Book1,SAW-Background-Book3}. That means, at comparable frequencies, their wavelengths differ by about five orders of magnitude. Thus, hybrid photon-phonon entanglement can provide a unique quantum correlation interface between macroscopic free-space photons and microscopic on-chip phonons, enabling the coherent exchange of spatial information across distinct spatial domains.

Entanglement provides the spatial correlation required for correlation imaging~\cite{Shih-Entangled-GI-1995,Wu-True-Thermal-2005,GI-Review1,GI-Review2,GI-Review3,GI-Book1}.~In most conventional studies, the probe field in the object arm and the readout field in the reference arm typically operate at identical or comparable wavelengths.~The probe wavelength should match the spatial features to be resolved~\cite{Imaging-Diffraction-Optics-Book,Imaging-Diffraction-Acoustics-Book}, whereas the readout wavelength need not be the same if suitable spatial correlations are established. Two-color correlation imaging has demonstrated this possibility using two optical fields of different wavelengths~\cite{Two-Color-Imaging-Chan,Two-Color-Imaging-Karmakar}. In practice, however, imaging usually faces the trade-off: the optimal field for probing an object is not suitable for readout due to environmental or operation constraints~\cite{Quantum-Sensing-Separation,Imaging-Separation-Security,Imaging-Separation-ElectronPhoton,Imaging-Separation-Biomedical}. This trade-off is particularly apparent in quantum radar~\cite{Macroscopic-MW-Detector-Cheng,Macroscopic-MW-Detector-Assouly}~and~on-chip defect inspection~\cite{Chip-SAW-Detector-Ohara1,Chip-SAW-Detector-Ohara2}: the former needs long-wavelength fields for probing macroscopic objects but compact and integratable readout devices, whereas the latter needs short-wavelength fields for resolving microscopic features could benefit from remote and non-invasive readout devices. These contrasting requirements call for the spatial correlation link distinct quantum fields, while preserving object's information across disparate spatial scales.

We here propose a quantum correlation imaging approach based on the hybrid entanglement between photons and phonons. By coupling a pump-driven superconducting quantum circuit simultaneously to a transmission line antenna for photons and an interdigital transducer (IDT) for phonons, we can generate a hybrid photon-phonon entanglement resource at comparable frequencies but with five order-of-magnitude wavelength disparity. This disparity circumvents the spatial-scale constraint in conventional correlation imaging, where the probe and readout fields typically operate at comparable wavelengths. We study two complementary imaging modalities: projecting photon-illuminated macroscopic objects onto microscopic phononic chips, and projecting phonon-illuminated microscopic objects into macroscopic photonic readout devices. This hybrid entanglement enables cross-scale spatial correlations between distinct quantum fields, allowing the probe field to be selected according to the object and sensing environment, independently of the readout field and readout device.

\begin{figure}[tb]
	\centering
	\includegraphics[width=8.6cm]{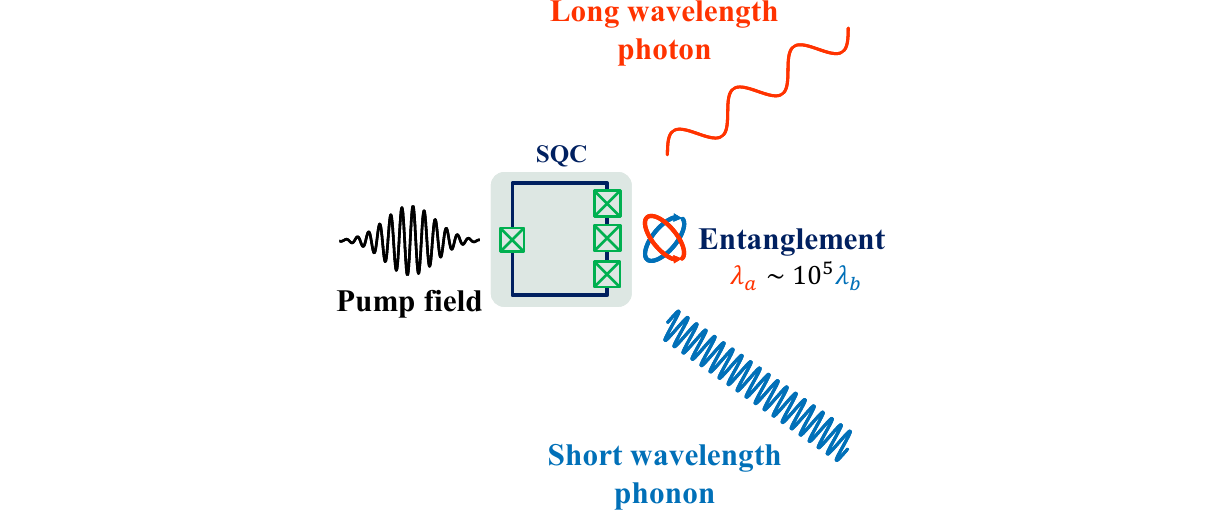}
	\caption{Conceptual schematic of correlation imaging via hybrid photon-phonon entanglement. A SQC, driven by a pump field, generates pairs of entangled photons and phonons at comparable frequencies with a wavelength disparity.}
	\label{F1ImagingConfiguration}
\end{figure}

{\it {Theoretical model and photon-phonon entanglement---}}The key to our imaging approach is the deterministic generation of hybrid entanglement between photons and phonons. As schematically shown in Fig.~\ref{F1ImagingConfiguration}, a superconducting quantum circuit (SQC), e.g., a SNAIL~\cite{SNAIL-Devoret1,SNAIL-Devoret2,SNAIL-Girvin} for concreteness, driven by a pump field of frequency $\omega_p$, can act as a nonlinear component~\cite{FluxQubit-Mooij,FluxQubit-Liu,SNAIL-Devoret1,SNAIL-Devoret2,SNAIL-Girvin} for mediating the generation of entangled photons and phonons via parametric conversion~\cite{Supplemental-Material}. When these two fields satisfy the frequency-matching condition $\omega_{a,j}+\omega_{b,j}=\omega_p$, the effective Hamiltonian in the interaction picture is given as~\cite{Supplemental-Material}
\begin{align}
H&=\sum_j
\big(G_j a_j^{\dagger} b_j^{\dagger}+
\text{H.c.}\big),
\label{Hamiltonian}
\end{align}
where the $j$th frequency-matched photon-phonon pair is described by photon $a_j^{\dagger}$ and phonon $ b_j^{\dagger}$ creation operators, corresponding to the frequencies $\omega_{a,j}$ and $\omega_{b,j}$, respectively, and $G_j$ is the effective photon-phonon coupling constant mediated by the SNAIL. Beginning with the vacuum state $\bigotimes_j |0_a,0_b\rangle_j$, one can generate entangled photon-phonon pairs via the Hamiltonian in Eq.~(\ref{Hamiltonian}). Under the weak pump approximation, the generated photon-phonon entangled state, subject to transverse momentum conservation~${q}_{a,j}+{q}_{b,j}=0$~\cite{Supplemental-Material}, is expressed in the continuous transverse wavevector basis~\cite{GI-Review1,GI-Review3,GI-Book1}~as
\begin{align}
|\psi\rangle&=\bigotimes_j
\iint dq_{a,j} dq_{b,j} \varphi_{ab,j}
a_j^{\dagger}(q_{a,j}) b_j^{\dagger}(q_{b,j})
|0_{a,j},0_{b,j}\rangle,
\label{EntanglementState}
\end{align}
where $a_j^{\dagger}(q_{a,j})$ and $b_j^{\dagger}(q_{b,j})$, obtained from the mode expansion of the operators $a_j^{\dagger}$ and $b_j^{\dagger}$~\cite{Supplemental-Material}, are the creation operators of a photon with transverse wavevector $q_{a,j}$ and a phonon with transverse wavevector $q_{b,j}$, respectively; $\varphi_{ab,j}\propto G_j$ denotes the joint probability amplitude for generating a photon and a phonon from the vacuum state.

The photon-phonon entangled state $|\psi\rangle$ exhibits a similar form to that used in conventional two-photon quantum correlation imaging~\cite{GI-Review1,GI-Review2,GI-Review3}. However, the frequency-matched photons and phonons differ in propagation velocity by about five orders of magnitude, leading to an extreme wavelength disparity, i.e., $\lambda_a \sim 10^5 \lambda_b$. This disparity fundamentally distinguishes the present hybrid entangled state from that for conventional correlation imaging. Hence, $|\psi\rangle$ emerges as a unique entanglement resource for cross-scale quantum sensing, and thus supports the two imaging modalities : the demagnification of photon-illuminated macroscopic objects into microscopic phononic chips, and conversely, the magnification of phonon-illuminated microscopic objects into macroscopic photonic readout devices.

The practical utility of the photon-phonon entangled state is determined by the generation efficiency of entangled photon-phonon pairs. This essential efficiency determines the available coincidence counts for reconstructing the object's spatial profile via correlation signals.~Here, we introduce the brightness~\cite{Brightness-Kuklewicz1,Brightness-Kuklewicz2,Brightness-Monteiro}~$B_{\rm pair}=R_{\text{pair}}/(P_{\text{pump}}\Delta\omega_{\text{eff}})$ to describe the generation efficiency of the desired entangled state in the long-time limit~\cite{Supplemental-Material}, which is dependent on the total generation rate $R_{\text{pair}}$ of frequency-matched photon-phonon pairs, the pump power $P_{\text{pump}}$, and the effective joint bandwidth $\Delta\omega_{\text{eff}}$ of generated photon-phonon pairs.~In correlation imaging, $B_{\text{pair}}$ should be large enough to extract the spatial information of the object. For typical circuit parameters~\cite{Supplemental-Material}, this brightness is estimated as $B_{\text{pair}} \simeq 3.9\times10^3$\,pairs/(s$\cdot$W$\cdot$Hz), ensuring the efficient measurement of correlation photon-photon pairs for imaging.

{\it {Correlation imaging via photon-phonon entanglement--}}
The spatial profile of a probed object is reconstructed through the spatial correlation between the photons and phonons at the detection planes.~The positive-frequency field operators at the transverse coordinate $x_{\mathcal{O},d}$ of the detection planes are given as~\cite{Supplemental-Material}
\begin{align}
E_{\text{obj},j}^{(+)}(x_{\mathcal{O},d})&=
\iint dx_{\mathcal{O},s} dx_{o} h_{\text{obj},so}^{(j)} T_o (x_o) h_{\text{obj},od}^{(j)} \mathcal{O}_j(x_{\mathcal{O},s})
\end{align}
for the object arm and
\begin{align}
E_{\text{ref},j}^{(+)}(x_{\mathcal{O},d})&=
\int dx_{\mathcal{O},s} h_{\text{ref},so}^{(j)} \mathcal{O}_j(x_{\mathcal{O},s})
\end{align}
for the reference arm. The subscripts obj and ref, labeling the object and reference arm, are assigned according to the imaging modalities: photons (phonons) serve as the object (reference) arm in the demagnification modality, i.e., obj$\rightarrow a$ and ref$\rightarrow b$, whereas their roles are exchanged in the magnification modality. Here, $\mathcal{O}_j(x_{\mathcal{O},s})=(2\pi)^{-1} \int d q_{\mathcal{O},j} \mathcal{O}_j (q_{\mathcal{O},j}) e^{i q_{\mathcal{O},j} x_{\mathcal{O},s}}$ represents the annihilation operator of mode $j$ for photon ($\mathcal{O}=a$) or phonon ($\mathcal{O}=b$) at the transverse coordinate $x_{\mathcal{O},s}$ of the source plane, and $T_o(x_o)$ denotes the object transmission function at the transverse coordinate $x_o$.~Hereafter, as a typical example, we consider a double-slit object described by~$T_o(x_o)=\text{rect}\left[(x_o-w_g)\big/w_s\right]+\text{rect}\left[(x_o+w_g)\big/w_s\right]$ with a slit width $w_s$ and slit distance $2w_g$. The propagation of these quantum fields is described by the function~\cite{Supplemental-Material} $h_{p,mn}=\text{Exp}\left[{i \pi (x_{p,m}-x_{p,n})^2}\big/{\lambda_{p} z_{p,mn}}\right]$, where $p \in \{\text{obj}, \text{ref}\}$, $z_{p,mn}$ is the longitudinal propagation distance, and $\{m,\,n\} \in \{s,\,o,\,d\}$ denote the source, object, and detection planes, respectively. Under the same modality-dependent mapping conversion, the propagation function automatically adopts the corresponding parameters $\{x_{p,m}, x_{p,n}, \lambda_{p}, z_{p,mn}\}$. These formulations provide a unified description of photon and phonon propagation dynamics in both imaging modalities.

The correlation function~$G^{(2)}(x_{a,d},x_{b,d})=\big\langle E_a^{(-)}(x_{a,d}) E_b^{(-)}(x_{b,d}) E_b^{(+)}(x_{b,d}) E_a^{(+)}(x_{a,d}) \big\rangle$ is obtained through photon-phonon coincidence measurements~\cite{Supplemental-Material}, from which the spatial profile of the probed object is reconstructed. Integration over the transverse coordinates of the spatially unresolved bucket detection plane yields~\cite{Supplemental-Material}
\begin{align}
G_{\bm{-}}^{(2)}(x_{b,d})&=
\int dx_{o} \left|T_o(x_o) K_{\bm{-}}(x_{b,d}, x_o)\right|^2, \label{eq:5}
\\
G_{\bm{+}}^{(2)}(x_{a,d})&=
\int dx_{o} \left|T_o(x_o) K_{\bm{+}}(x_{a,d}, x_o)\right|^2,
\label{Correlation-Function}
\end{align}	
where $K_{\bm{-}}(x_{b,d}, x_o)=\int dx_{b,s} h_{b,sd}(x_{b,s},x_{b,d}) h_{a,so}(x_{b,s},x_{o})$ and $K_{\bm{+}}(x_{a,d}, x_o)=\int dx_{b,s} h_{b,so}(x_{b,s},x_{o}) h_{a,sd}(x_{b,s},x_{a,d})$ denote the point-spread functions corresponding to the demagnification and magnification imaging modalities, respectively. These correlation function are further normalized by their peak values as ${I}_{\bm{\pm}}^{(2)}= {G_{\bm{\pm}}^{(2)}} \big/ {\text{Max}\big(G_{\bm{\pm}}^{(2)}\big)}$.

\begin{figure}[tb]
	\centering
	\includegraphics[width=8.6cm]{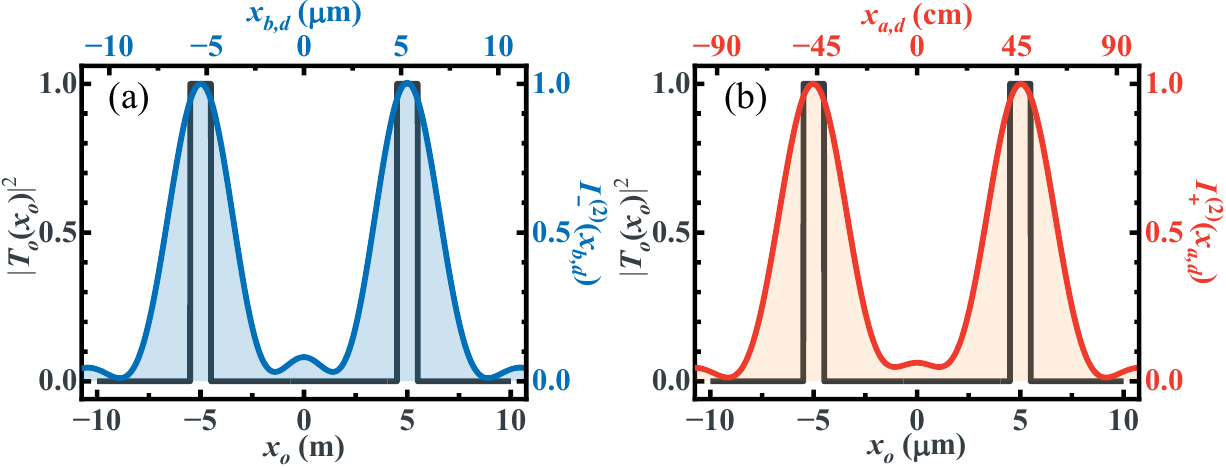}
	\caption{Cross-scale correlation images for: (a) the demagnification modality, and (b) the magnification modality. Blue and orange curves denote the normalized correlations~$I_{\bm{-}}^{(2)}(x_{b,d})$~and~$I_{\bm{+}}^{(2)}(x_{a,d})$, respectively, while gray curves represent the object transmission~$\left|T_o(x_o)\right|^2$.}
	\label{F2ImagingResults}
\end{figure}

The point-spread function $K_{\bm{\pm}}$, connecting the object plane $x_o$ with the spatially resolved detection planes ($x_{b,d}$ or $x_{a,d}$), fundamentally determines the projection of the object's spatial profile onto the measured correlation signal. In the ideal case of infinitely large source apertures, these point-spread functions are reduced to Dirac delta functions~\cite{Supplemental-Material}: $K_{\bm{-}} \rightarrow \delta(x_{b,d} + M_{\bm{-}} x_{o})$ for the demagnification modality, and $K_{\bm{+}} \rightarrow \delta(x_{a,d} + M_{\bm{+}} x_{o})$ for the magnification modality. Here, $M_{\bm{-}} = {\lambda_{b} z_{b,sd}} \big/ {\lambda_{a} z_{a,so}}$ and $M_{\bm{+}} = {\lambda_{a} z_{a,sd}} \big/ {\lambda_{b} z_{b,so}}$ denote the corresponding geometric scaling factors. Crucially, the large magnitude of these scaling factors arises from the combined effect of two distinct disparities on the spatial scale: the intrinsic five order-of-magnitude wavelength disparity $\lambda_a \sim 10^5 \lambda_b$ and the difference in longitudinal propagation distances between photons and phonons. Consequently, the correlation signal reconstructs the object's spatial profile scaled by a geometric factor, providing a giant demagnification $M_{\bm{-}}$ for compressing macroscopic free-space objects onto microscopic phononic chips, and conversely, a giant magnification $M_{\bm{+}}$ for expanding highly localized microscopic features into macroscopic photonic readout devices.

{\it {Imaging performance: spatial scaling, resolution and  visibility---}}
We now evaluate the correlation imaging performance in the realistic constraint of finite source apertures, wherein the point-spread functions $K_{\bm{\pm}}$ inevitably acquire finite spatial widths. This leads to the ultimate images being highly dependent on the system parameters, e.g., the source aperture, photonic and phononic wavelengths, and propagation distances. Despite this, the underlying cross-scale quantum correlation still enables the reconstruction of the object's spatial profile even with diffraction-induced broadening. As shown in Fig.~\ref{F2ImagingResults}(a), a photon-illuminated macroscopic object (e.g., a double-slit with a slit distance of about $10$\,m) described by $T_o(x_o)$ is projected onto the phononic chip as a highly demagnified image (with a slit-distance of about $10\,\SI{}{\micro\meter}$) described via $I_{\bm{-}}^{(2)}(x_{b,d})$. Conversely, a phonon-illuminated microscopic object (a double-slit with a slit-distance $10\,\SI{}{\micro\meter}$) described by $T_o(x_o)$ is projected into the photonic readout device as a highly magnified image (with a slit-distance of $90$\,cm) described via $I_{\bm{+}}^{(2)}(x_{a,d})$, as shown in Fig.~\ref{F2ImagingResults}(b).

\begin{figure}[tb]
	\centering
	\includegraphics[width=8.7cm]{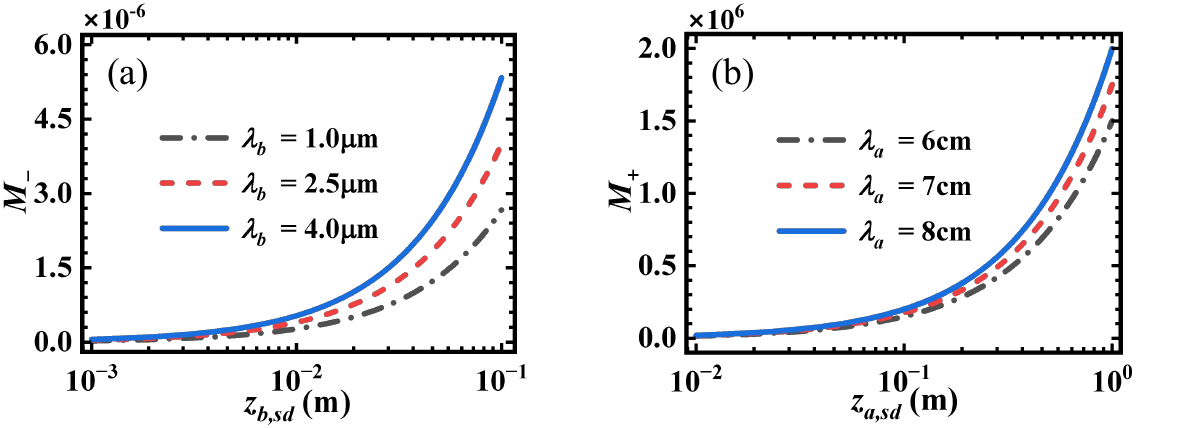}
	\caption{(a) Demagnification factor~vs.~longitudinal propagation distance $z_{b,sd}$ for different phononic wavelengths $\lambda_b$. (b) Magnification factor~vs.~longitudinal propagation distance $z_{a,sd}$ for different photonic wavelengths $\lambda_a$. Other parameters are the same as in Fig.~\ref{F2ImagingResults}.}
	\label{F3ImagingDemagnificationMagnification}
\end{figure}

As discussed above, the geometric scaling factors $M_{\bm{\pm}}$ depend highly on both the wavelength disparity between the fields in the object and reference arms, as well as the difference in their longitudinal propagation distances. In the demagnification modality, reducing either the phononic propagation distance $z_{b,sd}$ or phononic wavelength $\lambda_b$ decreases $M_{\bm{-}}$ as shown in Fig.~\ref{F3ImagingDemagnificationMagnification}(a), leading to a smaller image on phononic chip~\cite{Supplemental-Material}. Conversely, in the magnification modality, increasing either the photonic propagation distance $z_{a,sd}$ or photonic wavelength $\lambda_a$ increases $M_{\bm{+}}$ as shown in Fig.~\ref{F3ImagingDemagnificationMagnification}(b), resulting in a larger image in the photonic readout plane~\cite{Supplemental-Material}. The demagnification factor reaches $M_{\bm{-}} \simeq 2\times10^{-8}-5\times10^{-6}$ for typical parameters $\lambda_b \simeq 1-4\,\SI{}{\micro\meter}$ and $z_{b,sd} \simeq 1-100$\,mm, while the magnification factor reaches $M_{\bm{+}} \simeq 1\times10^{4}-2\times10^{6}$ for typical parameters $\lambda_a \simeq 6-8$\,cm, and $z_{a,sd} \simeq 1-100$\,cm. Importantly, these factors characterize also the spatial transformation between the object field of view and finite detector aperture, i.e., the reference-arm detector with effective aperture $D_{\rm ref}$ corresponding to the accessible field of view $D_{\rm obj}\simeq D_{\rm ref}/M_{\bm{\pm}}$. Thus, decreasing $M_{\bm{-}}$ expands the accessible view field of object for a fixed phononic chip size, while increasing $M_{\bm{+}}$ requires a larger photonic detector aperture. Additionally, since $\lambda_a$ and $\lambda_b$ are fixed by device design, their disparity establishes a static baseline of approximately $10^5$ for the scaling factor, leaving the propagation distance of the reference arm as a tunable parameter for matching the object size to the available readout platform.

\begin{figure}[tb]
	\centering
	\includegraphics[width=8.7cm]{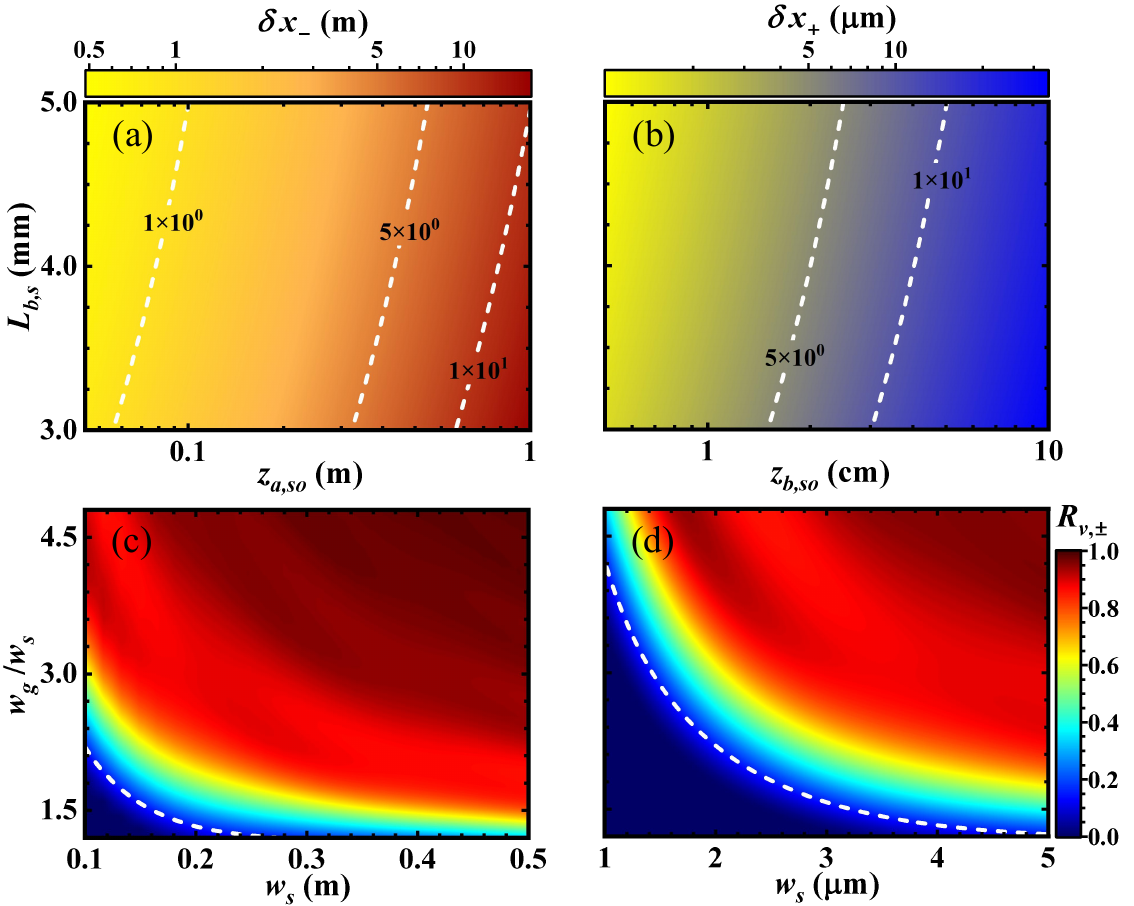}
	\caption{(a) Resolution of demagnification modality~vs.~phononic source aperture $L_{b,s}$ and photonic propagation distance $z_{a,so}$. (b) Resolution of magnification modality~vs.~phononic source aperture $L_{b,s}$ and phononic propagation distance $z_{b,so}$. (c) and (d): Imaging visibility~vs.~slit width $w_s$ and slit-distance $w_g$, corresponding to demagnification and magnification modalities in (a) and (b), respectively. The white dashed line marks $R_{v,\pm}\simeq0.15$, below which the correlation function signal becomes insufficient for reliable reconstruction of the object's spatial profile~\cite{Visibility-015-Howell}. Other parameters are the same as in Fig.~\ref{F2ImagingResults}.}
	\label{F4ImagingPerformance}
\end{figure}

Despite the giant spatial scaling, the resolution of both modalities is ultimately limited by the broadening of the point-spread function $K_{\pm}$.~The achievable resolutions are given by $\delta x_{\bm{+}}=\lambda_{b} z_{b,so} / L_{b,s}$ and $\delta x_{\bm{-}}=\lambda_{a} z_{a,so} / L_{b,s}$ for demagnification and magnification modalities, respectively.~That means, a larger phononic source aperture $L_{b,s}$, and a shorter wavelength or a shorter propagation distance of the probe field in the object arm lead to a narrower point-spread function and thus a sharper reconstructed image~\cite{Supplemental-Material}. Figures~\ref{F4ImagingPerformance}(a) and \ref{F4ImagingPerformance}(b) show the dependence of the resolutions $\delta x_{\bm{-}}$ and $\delta x_{\bm{+}}$ for the demagnification and magnification modalities on the propagation distance and phononic source aperture, respectively.~The demagnification resolution reaches $\delta x_{\bm{-}} \simeq 0.5-17$\,m for typical parameters $z_{a,so} \simeq 0.05-1$\,m and $L_{b,s} \simeq 3-5$\,mm, while the magnification resolution reaches $\delta x_{\bm{+}} \simeq 1-30\,\SI{}{\micro\meter}$ for typical parameters $z_{b,so} \simeq 0.5-10$\,cm and $L_{b,s} \simeq 3-5$\,mm. Additionally, a spatial sampling interval $\Delta x_{\rm ref}$ of the reference detector corresponds to an object-plane sampling interval $\Delta x_{\rm obj}=\Delta x_{\rm ref}/|M_{\bm{\pm}}|$. To ensure that sampling does not limit the achievable resolution $\delta x_{\bm{\pm}}$, one requires $\Delta x_{\rm obj} \lesssim \delta x_{\bm{\pm}}$, and thus $\Delta x_{\rm ref}/|M_{\bm{\pm}}| \lesssim \delta x_{\bm{\pm}}$.

While $\delta x_{\bm{\pm}}$ characterizes the resolution limit, the visibility of the correlation signal is also an important parameter that quantifies the image fidelity. For the double-slit, the visibility is defined as~$R_{v,\pm}=\big[\text{Max}\big(G_{\bm{\pm}}^{(2)}\big) - G_{\pm}^{(2)}(0)\big] \big/
\big[\text{Max}\big(G_{\bm{\pm}}^{(2)}\big) + G_{\pm}^{(2)}(0)\big]$, with the peak $\text{Max}\big(G_{\bm{\pm}}^{(2)}\big)$ and valley $G_{\pm}^{(2)}(0)$ correlation function values~\cite{Visibility-Mandel,Visibility-Born}.~Figures.~\ref{F4ImagingPerformance}(c) and \ref{F4ImagingPerformance}(d) show the visibility versus the double-slit parameters, for the typical parameters given in Figs.~\ref{F4ImagingPerformance}(a) and \ref{F4ImagingPerformance}(b).~Efficient reconstruction of the double-slit spatial profile requires the slit-distance to be larger than the slit width, which should be larger than the resolution limit. Moreover, the above analysis assumes the quasi-monochromatic approximation, wherein the point-spread function $K_{\pm}$ is frequency insensitive. Actually, a finite joint bandwidth of the generated photon-phonon pairs~\cite{Supplemental-Material} introduces frequency-dependent spatial correlation~\cite{GI-Review1}, since different modes of photons or phonons exhibit distinct propagation behavior.~The resulting multimode superposition would modify the resolution, visibility, and spatial scaling factors.~A sufficiently narrow joint bandwidth is therefore crucial for realizing high-fidelity correlation imaging.

{\it {Discussions and experimental feasibility---}}
Taking the photon-phonon entangled state as an example, we show that the hybrid quantum entanglement provides a unique resource for cross-scale correlation imaging with giant geometric scaling. However, the ultimate imaging performance remains limited by the probe wavelength and the phononic source aperture. An alternative way to circumvent this limitation is to go beyond the simple photon-phonon entangled state considered here and employ more sophisticated quantum states~\cite{Quantum-State-Imaging,NOON-Imaging,Fock-Imaging} to further improve the resolution. For instance, NOON states~\cite{NOON-Imaging}~exhibit $N$-fold phase accumulation and thus allow sub-wavelength spatial feature extraction for object profiles, whereas Fock states~\cite{Fock-Imaging}~provide number-state resources and thus improve visibility via higher-order conditional correlations.~Crucially, although a SNAIL is taken here as an example for generating the desired photon-phonon entangled state, the flexibility and tunability of SQCs~\cite{Hybrid-RMP-Xiang,Superconducting-Circuit1,Superconducting-Circuit2}, along with their simultaneous coupling to photons and phonons~\cite{QuantumAcoustics-Article1,SAW-Cavity-Multimode-TunableCoupling}, provides a platform for synthesizing a broader class of hybrid quantum states. This opens a route toward advanced correlation imaging with improved imaging performance.

Our correlation imaging approach based on hybrid quantum entanglement can be widely applied to other hybrid quantum architectures that can bond between distinct quantum fields, such as electric-optical transducers connecting microwave photons with optical photons~\cite{MW-Optical-Transducer-Sahu,MW-Optical-Transducer-Xu}, or acousto-optics transducers connecting surface acoustic wave phonons with optical photons~\cite{SAW-Optical-Transducer-Okada,SAW-Optical-Transducer-Lyer}. These architectures enable also the generation of hybrid entanglement resources that allow the optimal field for probing an object to be distinct from the one suited for readout. In this way, depending on the size of the object and operation environment, one can choose diverse quantum fields for probe and readout, including microwave photons~\cite{MW-Appl-Features}, surface acoustic wave phonons~\cite{SAW-Appl-Features},~and optical photons~\cite{Optical-Appl-Features}.

The realization of the present approach relies on the extraction of the spatial cross-correlation signal between entangled photons and phonons. At temperatures (about $10$-$20$\,mK) of a commercial dilution refrigerator~\cite{Dilution-Refrigerators} for solid-state quantum devices, the thermal occupation of both photonic and phononic modes at gigahertz-frequency can be suppressed, ensuring a sufficiently high generation efficiency of photon-phonon entangled states. Importantly, the essential transverse momentum conservation between frequency-matched photons and phonons can be realized by coupling SQC with an antenna for photons and a slanted IDT for phonons~\cite{Supplemental-Material}. The measurement of correlation signals requires time-resolved photon-phonon coincidences~\cite{Supplemental-Material}. Single photons can be measured via current-biased Josephson-junction detectors~\cite{MW-SinglePhotonCounter-Chen,MW-Correlation-Forgues}, whereas single phonons can be read out dispersively via the phonon-induced frequency shift of superconducting qubits~\cite{SAW-Qcontrol-Satzinger,SAW-Qcontrol-Dumur,SAW-SinglePhotonCounter-Qiao}. Accurate correlation signals require high-precision spatial sampling of the reference detector~\cite{Supplemental-Material}. This requirement is expected to be fulfilled with existing technologies, e.g., by integrating a cantilever-tip detector~\cite{SAW-Scanning-Detector-Tip,SAW-Scanning-Detector-Cantilever} with high-precision nano-positioning stages~\cite{SAW-Scanning-Detector-MotionTable} for on-chip phonon measurement, and by employing spatially resolved antenna arrays~\cite{MW-Array-Detector-Hansen,MW-Array-Detector-Nepa,MW-Array-Detector-Shao} for free-space photon measurement.

{\it {Conclusion---}}
In conclusion, we have proposed a correlation imaging approach based on the hybrid entanglement between photons and phonons. Due to the wavelength disparity between the two quantum fields, this entanglement resource allows spatial information to be transferred across vastly different spatial scales. Consequently, macroscopic free-space object profiles can be demagnified onto microscopic phononic chips, and conversely, on-chip microscopic object profiles can be magnified onto macroscopic photonic readout devices. Our study provides a route toward hybrid-field quantum sensing, where the entangled resource can be engineered across different quantum systems and chosen independently to match the object scale and operation environment. These results may open a new avenue for quantum correlation imaging and cross-scale quantum information processing.

\begin{acknowledgments}
This work was supported by the National Natural Science Foundation of China with Grants No.~92365209, No.~12374483 and No.~62474012.
\end{acknowledgments}

\end{document}